\documentclass[amsmath,amssymb,11pt]{revtex4}

\usepackage{epsfig,amssymb}
\usepackage{graphicx}
\usepackage{dcolumn}
\usepackage{bm}
\usepackage{feynmp}
\def\r{{\bf r}}
\def\rp{{\bf r^\prime}}
\def\rpp{{\bf r^{\prime\prime}}}
\def\G{{\bf G}}
\def\Gp{{\bf G^{\prime}}}
\def\q{{\bf q}}
\def\k{{\bf k}}
\def\R{{\bf R}}
\begin{document}
\begin{fmffile}{graph2}
\setlength{\unitlength}{1mm}

\title{ Embedding theory for excited states with inclusion of
  self-consistent environment screening}

\author{Johannes Lischner}\address{Laboratory of Atomic and Solid State Physics, Cornell University, Ithaca, New York 14853}

\author{Dominika Zgid}\address{Department of Chemistry and Chemical Biology, Cornell University, Ithaca, New York 14853}

\author{Garnet Kin-Lic Chan}\address{Department of Chemistry and
  Chemical Biology, Cornell University, Ithaca, New York 14853}

\author{T.A. Arias}\address{Laboratory of Atomic and Solid State Physics, Cornell University, Ithaca, New York 14853}

\begin{abstract}
We present a general embedding theory of electronic excitations of a
relatively small, localized system in contact with an extended,
chemically complex environment. We demonstrate how to include the
screening response of the environment into highly accurate electronic
structure calculation of the localized system by means of an effective
interaction between the electrons, which contains \emph{only screening
  processes occurring in the environment}. For the common case of a
localized system which constitutes an inhomogeneity in an otherwise
homogeneous system, such as a defect in a crystal, we show how matrix
elements of the environment-screened interaction can be calculated
from density-functional calculations of the \emph{homogeneous} system
\emph{only}.  We apply our embedding theory to the calculation of
excitation energies in crystalline ethylene.

\end{abstract}

\maketitle

\section{Introduction}
Neutral electronic excited states play an important role in the study
of condensed matter systems. They are probed in spectroscopic
measurements, such as absorption, reflectivity or
photoluminescence. In addition, they are important in many technical
applications, such as photovoltaics, laser technology or
light-emitting diodes.  It is therefore necessary to develop a
theoretical understanding of neutral electronic excitations.

Various theoretical approaches to calculating excited state properties
have been developed in the past, each of which usually is applicable
to certain classes of physical systems. Highly accurate quantum
chemistry methods, such as configuration interaction or coupled
cluster theory, can only be applied to small systems containing few
electrons, such as atoms or small molecules. For larger molecules and
clusters, time-dependent density-functional theory
\cite{RungeGross,RubioPhysRevMod} yields a reliable description of
excited states. Many-body perturbation theory, where neutral
excitation energies and oscillator strengths are obtained by solving
the Bethe-Salpeter equation for the two-particle Green function,
constitutes a standard tool for neutral excited states in extended
periodic systems, but has also been applied successfully to localized
systems, such as molecules
\cite{LouieRohlfing,LouieGrossmann,Bechstedt}. However, many-body
perturbation theory, while being applicable to much larger systems
than the quantum chemistry methods, still scales quite unfavorably
with system size and often further approximation, such as model
dielectric functions, are introduced when studying chemically complex
crystals \cite{Bobbert1,Bobbert2}.

Often, in extended systems one can identify a subsystem, which is of
particular interest: for example, an adsorbed molecule near a solid
surface or a defect in a crystal. In such systems, the electronic
excitation is often localized on the special subsystem and a high
level of accuracy is desired for the description of the excitation.
However, while standard approaches for large systems, such as
time-dependent density-functional theory, often do not yield the
required accuracy, highly accurate quantum chemistry approaches can
only be employed when the environment, which screens the potential
created by the localized excitation, is ignored.  Embedding theories,
which attempt to combine a high-accuracy treatment of the subsystem
with a more approximate treatment of the environment, attempt to
overcome this difficulty. Those theories have a long history, but have
focused mostly on ground-state properties \cite{CarterReview}.

Regarding excited state embedding, Whitten and coworkers
\cite{Whitten} construct an embedding scheme, where a configuration
interaction calculation of the subsystem is based on a Hartree-Fock
calculation of a surrounding cluster. However, the description of the
environment by a cluster is a severe approximation, which neglects
long range screening effects. Carter and coworkers
\cite{CarterHuang,GovindCarter1,GovindCarter2,KlunerCarter1,KlunerCarter2}
building on previous work\cite{Cortona,Wesolowski} embed a
configuration interaction calculation into an environment described by
density-functional theory. Starting from a formally exact set of
equations, Carter and coworkers \cite{CarterHuang} compute excitation
energies of the subsystem by introducing an additional ``external''
potential due to the environment into their quantum chemistry
calculation. However, they assume that the electron density of the
environment in the excited state remains the same as in the ground
state, neglecting the rearrangement of the environment electrons due
to the excitation of the subsystem.

In this work, we present a theory of electronic excitations of a
localized system in contact with an extended, chemically complex
environment taking proper account of the environment response to the
excitation on the localized system. In particular, we include the
environment response through a special screened interaction, which
contains only screening processes of the environment, acting between
electrons of the subsystem. This environment-screened interaction is
obtained by first carrying out density-functional calculations of the
\emph{homogeneous} system, e.g. the defect-free crystal, and then
subtracting out screening processes due to the region which is
replaced by the explicit system (e.g. defect).

Once obtained, the environment-screened interaction can be used in
various electronic structure methods for the localized, ``explicit''
system: we demonstrate how to include the environment response into a
Green function calculation of the subsystem or into a wavefunction
calculation.

This paper is organized as follows: in Section 1, we introduce the
concept of an environment-screened interaction and describe its use in
an embedding procedure, where the explicit region is treated by Green
function methods. In Section 2, we describe the application of our
embedding approach to wavefunctions methods, which are typically
employed in quantum chemistry calculations. In particular, we
introduce a self-consistent equation for the environment-screened
potential caused by an external charge distribution and consequently
transform this equation into a matrix form. We also discuss numerical
difficulties of our embedding approach which are caused by (i) the
divergence of the screened interaction and (ii) the extreme narrowness
of the basis functions used in quantum chemistry. The first problem is
solved by introducing a new integration scheme for integrals over the
Brillouin zone, which avoids the approximations made in similar
approaches. The second difficulty is overcome by computing the
\emph{changes} of the interaction matrix elements due to the presence
of the environment, which can be evaluated efficiently in Fourier
space because only the smooth parts of the basis functions are
screened. In Section 3, we apply our theory to crystalline ethylene
and discuss the effects of the crystalline environment on the excited
states of these materials. Section 4 offers discussions and outlook.

\section{Embedding theory and application Green function methods}

The identification of a localized subsystem, on which the modeling
effort is concentrated, marks the starting point in every embedding
approach. In our embedding approach for excited states, the physical
system under consideration is divided into two regions: a localized
``explicit'' region, where the excitation occurs, and the rest of the
system, the environment, which screens the potential created by the
excitation. Having identified the relevant subsystem, we employ highly
accurate, but computationally demanding electronic structure methods,
such as the Green function methods or wavefunction methods, to
describe the localized excitation. To include the influence of the
environment into these calculations, we modify the interaction between
the electrons to contain all screening processes occurring in the
environment. This environment-screened interaction is obtained from
less demanding many-body Green's function methods, evaluated
eventually with the Kohn-Sham orbitals from density-functional theory
calculations.

\subsection{Environment-screened interaction}

The environment-screened interaction between electrons of the
localized system, where the excitation occurs, plays a crucial role in
our embedding approach. Quite often, it is possible and advantageous
to replace a complicated calculation of a self-consistent environment
response by the use of an effective or screened interaction, the
simplest case being the interaction of two point charges in a linear
dielectric medium. The notion of a screened interaction also plays an
important role in quantum many-body theory, where Feynman diagrams are
used to organize and gain intuition about the plethora of possible
screening processes. Here we also employ a diagrammatic approach to
derive an explicit expressions for the environment-screened
interaction.

The rules for converting Feynman diagrams into algebraic expressions
can be found in many textbooks,
e.g. Ref.~\cite{Mahan,FetterWalecka}. In the diagrams below, dotted
lines denote bare Coulomb interactions, while the fully screened
(containing screening processes from the environment \emph{and} the
localized subsystem) interaction is represented by double-dotted
lines. As usual, electron (and holes) are denoted by regular
lines. The screened interaction is given by
\begin{equation}
  \parbox{20mm}{\begin{fmfgraph}(20,10)
      \fmfleft{i1} \fmfright{o1}
      \fmf{dbl_dashes}{i1,o1}
  \end{fmfgraph}}
  \;=\;
  \parbox{15mm}{\begin{fmfgraph}(15,10)
      \fmfleft{i1} \fmfright{o1}
      \fmf{dashes}{i1,o1}
  \end{fmfgraph}}  
  \;+\;
  \parbox{25mm}{\begin{fmfgraph}(25,20)
      \fmfleft{i1} \fmfright{o1}
      \fmf{dashes}{i1,v1} \fmf{dashes}{v2,o1}
      \fmf{plain,left,tension=0.7}{v1,v2,v1}
  \end{fmfgraph}}
  \;+\;
  \parbox{40mm}{\begin{fmfgraph}(40,20)
      \fmfleft{i1} \fmfright{o1}
      \fmf{dashes}{i1,v1} \fmf{dashes}{v2,v3}
      \fmf{plain,left,tension=0.7}{v1,v2,v1}
      \fmf{plain,left,tension=0.7}{v3,v4,v3}
      \fmf{dashes}{v4,o1}
  \end{fmfgraph}}
  \;+\;
  \parbox{30mm}{\begin{fmfgraph}(30,20)
      \fmfleft{i1} \fmfright{o1}
      \fmftop{i2}  \fmfbottom{o2}
      \fmf{phantom,tension=1}{i2,v2}
      \fmf{phantom,tension=1}{v4,o2}
      \fmf{dashes,tension=2}{i1,v1}
      \fmf{plain,left=0.3,tension=1}{v1,v2,v3,v4,v1}
      \fmf{dashes,tension=2}{v3,o1}
      \fmf{dashes,tension=0}{v2,v4}
  \end{fmfgraph}}
  \;+\; ...
  \label{ScreenedInteraction}
\end{equation}

By introducing the irreducible polarizability, represented below by a
dashed square, we can express the infinite sum by a self-consistent
Dyson equation. The irreducible polarizability contains all diagrams
which cannot be separated into two parts by cutting a single dotted
line. Dyson's equation for the screened interaction is given by

\begin{equation}
 \parbox{30mm}{\begin{fmfgraph}(30,20)
      \fmfleft{i1} \fmfright{o1}
      \fmf{dbl_dashes}{i1,o1}
  \end{fmfgraph}}
\;=\;
 \parbox{30mm}{\begin{fmfgraph}(30,20)
      \fmfleft{i1} \fmfright{o1}
      \fmf{dashes}{i1,o1}
 \end{fmfgraph}}
\;+\;
 \parbox{40mm}{\begin{fmfgraph}(40,20)
      \fmfleft{i1} \fmfright{o1}
      \fmf{dashes}{i1,v1}
      \fmf{dbl_dashes}{v1,o1}
      \fmfv{decor.shape=square,tension=0.5,decor.filled=shaded}{v1}
 \end{fmfgraph}}
 \label{Dyson}.
\end{equation}

Of course, the exact form of the irreducible polarizability is unknown
and must be approximated. A well-known approximation, which has proven
extremely useful in many applications of electronic structure theory,
is the random-phase approximation (RPA), where only simple
electron-hole bubbles are inserted into the dotted lines in
Eq.~(\ref{ScreenedInteraction}).

The electron and hole of each bubble are either part of the explicit
subsystem or part of the environment. In the following, Green function
lines corresponding to subsystem electrons (and holes) are colored
red, environment electrons (and holes) are colored blue. Consequently,
a diagram representing an electron-hole bubble gives rise to four
colored diagrams

\begin{equation}
  \parbox{27mm}{\begin{fmfgraph}(27,20)
      \fmfleft{i1} \fmfright{o1}
      \fmf{dashes}{i1,v1} \fmf{dashes}{v2,o1}
      \fmf{plain,left}{v1,v2} \fmf{plain,left}{v2,v1}
  \end{fmfgraph}}
  \;\; \rightarrow \;\;
  \parbox{27mm}{\begin{fmfgraph}(27,20)
      \fmfleft{i1} \fmfright{o1}
      \fmf{dashes}{i1,v1} \fmf{dashes}{v2,o1}
      \fmf{plain,left,foreground=blue}{v1,v2} \fmf{plain,left,foreground=blue}{v2,v1}
  \end{fmfgraph}}
  \; , \;
  \parbox{27mm}{\begin{fmfgraph}(27,20)
      \fmfleft{i1} \fmfright{o1}
      \fmf{dashes}{i1,v1} \fmf{dashes}{v2,o1}
      \fmf{plain,left,foreground=red}{v1,v2} \fmf{plain,left,foreground=red}{v2,v1}
  \end{fmfgraph}}
  \; , \;
  \parbox{27mm}{\begin{fmfgraph}(27,20)
      \fmfleft{i1} \fmfright{o1}
      \fmf{dashes}{i1,v1} \fmf{dashes}{v2,o1}
      \fmf{plain,left,foreground=blue}{v1,v2} \fmf{plain,left,foreground=red}{v2,v1}
  \end{fmfgraph}}
  \; , \;
  \parbox{27mm}{\begin{fmfgraph}(27,20)
      \fmfleft{i1} \fmfright{o1}
      \fmf{dashes}{i1,v1} \fmf{dashes}{v2,o1}
      \fmf{plain,left,foreground=red}{v1,v2} \fmf{plain,left,foreground=blue}{v2,v1}
  \end{fmfgraph}} \; .
\end{equation}

In the first colored Feynman diagram an electron-hole pair is created
in the environment, while in the second diagram the pair is created in
the explicit region. In the third and fourth diagrams one particle is
created in the environment, while the other is created in the explicit
system. The contribution to the total screening from such diagrams,
where environment particle lines and system particle lines connect, is
small if there is little wavefunction overlap between subsystem and
environment. Algebraically, such an electron-hole bubble corresponds
to the product of a Green function $G_{sys}(r,r')$ describing
particles in the explicit system and a Green function $G_{env}(r,r')$
describing environment particles. Because the bare Green function is
proportional $\psi(r)\psi^*(r')$, the product will vanish if
environment and system wavefunctions do not overlap.

Within the RPA, the fully screened interaction can therefore be
expressed as
\begin{equation}
\parbox{15mm}{\begin{fmfgraph}(15,20)
      \fmfleft{i1} \fmfright{o1}
      \fmf{dbl_dashes}{i1,o1}
  \end{fmfgraph}}
\;= \;
 \parbox{15mm}{\begin{fmfgraph}(15,20)
      \fmfleft{i1} \fmfright{o1}
      \fmf{dashes}{i1,o1}
 \end{fmfgraph}}
\; + \;
\parbox{23mm}{\begin{fmfgraph}(23,20)
      \fmfleft{i1} \fmfright{o1}
      \fmf{dashes}{i1,v1} \fmf{dashes}{v2,o1}
      \fmf{plain,left,foreground=red,tension=0.7}{v1,v2,v1}
      \end{fmfgraph}}
\; + \;
\parbox{23mm}{\begin{fmfgraph}(23,20)
      \fmfleft{i1} \fmfright{o1}
      \fmf{dashes}{i1,v1} \fmf{dashes}{v2,o1}
      \fmf{plain,left,foreground=blue,tension=0.7}{v1,v2,v1}
 \end{fmfgraph}}
\; + \;
\parbox{28mm}{\begin{fmfgraph}(28,20)
      \fmfleft{i1} \fmfright{o1}
      \fmf{dashes}{i1,v1} \fmf{dashes}{v2,v3} \fmf{dashes}{v4,o1}
      \fmf{plain,left,foreground=blue,tension=0.5}{v1,v2,v1}
      \fmf{plain,left,foreground=red,tension=0.5}{v3,v4,v3}
 \end{fmfgraph}}
\;+ \;
\parbox{28mm}{\begin{fmfgraph}(28,20)
      \fmfleft{i1} \fmfright{o1}
      \fmf{dashes}{i1,v1} \fmf{dashes}{v2,v3} \fmf{dashes}{v4,o1}
      \fmf{plain,left,foreground=blue,tension=0.5}{v1,v2,v1}
      \fmf{plain,left,foreground=blue,tension=0.5}{v3,v4,v3}
 \end{fmfgraph}}
\;+ \;...
\end{equation}

We now define the environment-screened interaction, which is
represented by a blue double-dotted line and contains only environment
bubbles, according to

\begin{equation}
\parbox{15mm}{\begin{fmfgraph}(15,20)
      \fmfleft{i1} \fmfright{o1}
      \fmf{dbl_dashes,foreground=blue}{i1,o1}
  \end{fmfgraph}}
\;=\;
 \parbox{15mm}{\begin{fmfgraph}(15,20)
      \fmfleft{i1} \fmfright{o1}
      \fmf{dashes}{i1,o1}
 \end{fmfgraph}}
\;+\;
\parbox{23mm}{\begin{fmfgraph}(23,20)
      \fmfleft{i1} \fmfright{o1}
      \fmf{dashes}{i1,v1} \fmf{dashes}{v2,o1}
      \fmf{plain,left,foreground=blue,tension=0.7}{v1,v2,v1}
      \end{fmfgraph}}
\;+\;
\parbox{30mm}{\begin{fmfgraph}(30,20)
      \fmfleft{i1} \fmfright{o1}
      \fmf{dashes}{i1,v1} \fmf{dashes}{v2,v3} \fmf{dashes}{v4,o1}
      \fmf{plain,left,foreground=blue,tension=0.5}{v1,v2,v1}
      \fmf{plain,left,foreground=blue,tension=0.5}{v3,v4,v3}
 \end{fmfgraph}}
\;+\;
\parbox{35mm}{\begin{fmfgraph}(35,20)
      \fmfleft{i1} \fmfright{o1}
      \fmf{dashes}{i1,v1} \fmf{dashes}{v2,v3} \fmf{dashes}{v4,v5}
      \fmf{dashes}{v6,o1}
      \fmf{plain,left,foreground=blue,tension=0.5}{v1,v2,v1}
      \fmf{plain,left,foreground=blue,tension=0.5}{v3,v4,v3}
      \fmf{plain,left,foreground=blue,tension=0.5}{v5,v6,v5}
\end{fmfgraph}}
\;+\; ...
\end{equation}

Note that the fully (within the RPA) screened interaction can be
obtained from the environment-screened one via
\begin{equation}
\parbox{15mm}{\begin{fmfgraph}(15,20)
      \fmfleft{i1} \fmfright{o1}
      \fmf{dbl_dashes}{i1,o1}
  \end{fmfgraph}}
\;=\;
 \parbox{15mm}{\begin{fmfgraph}(15,20)
      \fmfleft{i1} \fmfright{o1}
      \fmf{dbl_dashes,foreground=blue}{i1,o1}
 \end{fmfgraph}}
\;+\;
\parbox{25mm}{\begin{fmfgraph}(25,20)
      \fmfleft{i1} \fmfright{o1}
      \fmf{dbl_dashes,foreground=blue}{i1,v1}
      \fmf{dbl_dashes,foreground=blue}{v2,o1}
      \fmf{plain,left,foreground=red,tension=0.7}{v1,v2,v1}
      \end{fmfgraph}}
\;+\;
\parbox{30mm}{\begin{fmfgraph}(30,20)
      \fmfleft{i1} \fmfright{o1}
      \fmf{dbl_dashes,foreground=blue}{i1,v1}
      \fmf{dbl_dashes,foreground=blue}{v2,v3}
      \fmf{dbl_dashes,foreground=blue}{v4,o1}
      \fmf{plain,left,foreground=red,tension=0.5}{v1,v2,v1}
      \fmf{plain,left,foreground=red,tension=0.5}{v3,v4,v3}
 \end{fmfgraph}}
\;+\;
\parbox{40mm}{\begin{fmfgraph}(40,20)
      \fmfleft{i1} \fmfright{o1}
      \fmf{dbl_dashes,foreground=blue}{i1,v1}
      \fmf{dbl_dashes,foreground=blue}{v2,v3}
      \fmf{dbl_dashes,foreground=blue}{v4,v5}
      \fmf{dbl_dashes,foreground=blue}{v6,o1}
      \fmf{plain,left,foreground=red,tension=0.5}{v1,v2,v1}
      \fmf{plain,left,foreground=red,tension=0.5}{v3,v4,v3}
      \fmf{plain,left,foreground=red,tension=0.5}{v5,v6,v5}
\end{fmfgraph}}
\;+\; ...
\label{RPAscreened}
\end{equation}

This simple result has important consequences: it allows us to
rigorously replace the whole system, consisting of the localized
subsystem and the environment, by an ``effective'' subsystem, in which
electrons interact through the environment-screened interaction.

We note that this result is not limited to the random-phase
approximation of the screened interaction, but is found quite
generally whenever the irreducible polarizability is a sum of two
pieces: one containing only environment electrons, the other only
system electrons. This opens up the possibility of concentrating the
modelling effort on the localized subsystem by using more
sophisticated approaches for this region than the RPA.  In our
approach, we employ the RPA for the environment-screened interactions
and use highly accurate electronic structure methods, such as Green
function theory or explicit wavefunction approaches, for the localized
subsystem.

\subsection{Embedding in Green function methods}
In this section, we demonstrate how to use the environment-screened
interaction to construct an embedding theory, where the excitation on
the localized system is described by Green function methods.

The Bethe-Salpeter equation of many-body perturbation theory solves
for the dressed particle-hole Green function
\cite{LouieRohlfing,RubioPhysRevMod}, whose poles give the excitation
energies of the system, and has been applied successfully to extended
periodic systems, but also clusters and molecules
\cite{LouieRohlfing,LouieGrossmann}.

Diagrammatically, the Bethe-Salpeter equation is given by
\begin{equation}
  \parbox{20mm}{\begin{fmfgraph}(20,15)
      \fmfleft{i1,i2} \fmfright{o1,o2}
      \fmf{plain}{i1,v2} \fmf{plain}{i2,v1}
      \fmfpoly{smooth,tension=0.5,filled=shaded}{v1,v2,v3,v4}
      \fmf{plain}{v3,o1} \fmf{plain}{v4,o2}
  \end{fmfgraph}}
  =
  \parbox{20mm}{\begin{fmfgraph}(20,15)
      \fmfleft{i1,i2} \fmfright{o1,o2}
      \fmf{plain}{i1,o1} \fmf{plain}{i2,o2}
  \end{fmfgraph}}
  +
  \parbox{20mm}{\begin{fmfgraph}(20,15)
      \fmfleft{i1,i2} \fmfright{o1,o2}
      \fmf{plain}{i1,o1} \fmf{plain}{i2,o2}
  \end{fmfgraph}}
  \left(
  \parbox{20mm}{\begin{fmfgraph}(20,15)
      \fmfleft{i1,i2} \fmfright{o1,o2}
      \fmf{plain}{i1,v1} \fmf{plain}{i2,v1}
      \fmf{plain}{v2,o1} \fmf{plain}{v2,o2}
      \fmf{dashes}{v1,v2}
  \end{fmfgraph}}
  +
  \parbox{20mm}{\begin{fmfgraph}(20,15)
      \fmfleft{i1,i2} \fmfright{o1,o2}
      \fmf{plain}{i1,v1} \fmf{plain}{i2,v2}
      \fmf{plain}{v1,o1} \fmf{plain}{v2,o2}
      \fmf{dbl_dashes}{v1,v2}
  \end{fmfgraph}}
  \right)
  \parbox{20mm}{\begin{fmfgraph}(20,15)
      \fmfleft{i1,i2} \fmfright{o1,o2}
      \fmf{plain}{i1,v2} \fmf{plain}{i2,v1}
      \fmfpoly{smooth,tension=0.5,filled=shaded}{v1,v2,v3,v4}
      \fmf{plain}{v3,o1} \fmf{plain}{v4,o2}
  \end{fmfgraph}},
  \label{BetheSalpeter}
\end{equation}

where the term in parenthesis represents the Bethe-Salpeter
approximation to the self-energy of the particle-hole Green function
and contains an unscreened exchange interaction and a fully screened
direct interaction. Typically, the fully screened interaction is
obtained from the static dielectric matrix within the random-phase
approximation \cite{LouieRohlfing,RubioPhysRevMod}.

To obtain an embedding theory within the Bethe-Salpeter framework, we
again color the electron lines red for system electrons and blue for
environment electrons. Assuming that the exciton is localized in the
explicit region, we impose that the incoming and outgoing lines in
Eq.~(\ref{BetheSalpeter}) are red system lines. The direct interaction
already is fully screened containing the screening contributions from
both the localized system and the environment. However, expressing the
self-consistent Bethe-Salpeter equation (Eq.~(\ref{BetheSalpeter})) by
an infinite sum of diagrams, we observe that the exchange interaction
in the two-particle self-energy gives rise to diagrams like
\begin{equation}
\parbox{30mm}{\begin{fmfgraph}(30,20)
      \fmfleft{i1,i2} \fmfright{o1,o2}
      \fmf{dashes,tension=2}{v1,v2}
      \fmf{plain,left=1.2,foreground=blue}{v2,v3,v2}
      \fmf{dashes,tension=2}{v3,v4}
      \fmf{plain,foreground=red,tension=2}{i1,v1}
      \fmf{plain,foreground=red,tension=2}{i2,v1}
      \fmf{plain,foreground=red,tension=2}{v4,o1}
      \fmf{plain,foreground=red,tension=2}{v4,o2}
  \end{fmfgraph}}.
\label{EnvExch}
\end{equation}
To capture such an exchange-type process, where a particle-hole pair
of the localized system annihilates and re-emerges after creating an
intermediate electron-hole pair in the environment, in our embedding
approach, we replace the bare exchange diagram in
Eq.~(\ref{BetheSalpeter}) by an environment-screened exchange.

It is easy to verify that the following Bethe-Salpeter equation for
the embedded system
\begin{equation}
  \parbox{20mm}{\begin{fmfgraph}(20,15)
      \fmfleft{i1,i2} \fmfright{o1,o2}
      \fmf{plain,foreground=red}{i1,v2} \fmf{plain,foreground=red}{i2,v1}
      \fmfpoly{smooth,tension=0.5,filled=shaded}{v1,v2,v3,v4}
      \fmf{plain,foreground=red}{v3,o1} \fmf{plain,foreground=red}{v4,o2}
  \end{fmfgraph}}
  =
  \parbox{20mm}{\begin{fmfgraph}(20,15)
      \fmfleft{i1,i2} \fmfright{o1,o2}
      \fmf{plain,foreground=red}{i1,o1} \fmf{plain,foreground=red}{i2,o2}
  \end{fmfgraph}}
  +
  \parbox{20mm}{\begin{fmfgraph}(20,15)
      \fmfleft{i1,i2} \fmfright{o1,o2}
      \fmf{plain,foreground=red}{i1,o1} \fmf{plain,foreground=red}{i2,o2}
  \end{fmfgraph}}
  \left(
  \parbox{20mm}{\begin{fmfgraph}(20,15)
      \fmfleft{i1,i2} \fmfright{o1,o2}
      \fmf{plain,foreground=red}{i1,v1} \fmf{plain,foreground=red}{i2,v1}
      \fmf{plain,foreground=red}{v2,o1} \fmf{plain,foreground=red}{v2,o2}
      \fmf{dbl_dashes,foreground=blue}{v1,v2}
  \end{fmfgraph}}
  +
  \parbox{20mm}{\begin{fmfgraph}(20,15)
      \fmfleft{i1,i2} \fmfright{o1,o2}
      \fmf{plain,foreground=red}{i1,v1} \fmf{plain,foreground=red}{i2,v2}
      \fmf{plain,foreground=red}{v1,o1} \fmf{plain,foreground=red}{v2,o2}
      \fmf{dbl_dashes}{v1,v2}
  \end{fmfgraph}}
  \right)
  \parbox{20mm}{\begin{fmfgraph}(20,15)
      \fmfleft{i1,i2} \fmfright{o1,o2}
      \fmf{plain,foreground=red}{i1,v2} \fmf{plain,foreground=red}{i2,v1}
      \fmfpoly{smooth,tension=0.5,filled=shaded}{v1,v2,v3,v4}
      \fmf{plain,foreground=red}{v3,o1} \fmf{plain,foreground=red}{v4,o2}
  \end{fmfgraph}}
\label{BSEembed}
\end{equation}
indeed gives rise to diagrams like Eq.~(\ref{EnvExch}).

However, apart from the aforementioned embedding approximations
(neglect of diagrams with environment lines connecting into system
lines and the assumption that the exciton resides on the explicit
system), the proposed embedded Bethe-Salpeter equation contains one
further approximation: all environment bubbles are empty,
i.e. diagrams like
\begin{equation}
\parbox{35mm}{\begin{fmfgraph}(35,25)
      \fmfleft{i1,i2} \fmfright{o1,o2}
      \fmftop{i3} \fmfbottom{i4}
      \fmf{plain,foreground=red,tension=1}{i1,w1}
      \fmf{plain,foreground=red,tension=1}{i2,w1}
      \fmf{dashes,straight}{w1,v1}
      \fmf{phantom,tension=1}{i3,v2}
      \fmf{phantom,tension=1}{i4,v3}
      \fmf{plain,left=0.5,foreground=blue}{v1,v2,w2,v3,v1}
      \fmf{dbl_dashes,tension=0}{v2,v3}
      \fmf{dashes,straight}{w2,v5}
      \fmf{plain,foreground=red,tension=1}{v5,o1}
      \fmf{plain,foreground=red,tension=1}{v5,o2}
  \end{fmfgraph}},
\label{missing}
\end{equation}
which can be constructed from Eq.~(\ref{BetheSalpeter}), are absent in
Eq.~(\ref{BSEembed}). In these diagrams, additional interactions
between electron-hole pairs of the environment are included.
Nevertheless, we expect that Eq.~(\ref{BSEembed}) captures the vast
majority of all screening processes.

Computationally, the embedded Bethe-Salpeter equation has the
advantage that, once the environment-screened interaction is computed,
Eq.~(\ref{BSEembed}) can be conveniently solved in a localized basis,
such as gaussians \cite{LouieRohlfing}, with relatively few basis
functions.

\section{Embedding in wavefunction methods}

Having introduced the general notion of an environment-screened
interaction and having seen its use in an embedded Bethe-Salpeter
equation, we now move on to embed a localized system described by
wavefunction methods, which are typically used in quantum chemistry
calculations, into an environment described by the random phase
approximation of many body theory.

We use the Lehmann representation of the particle-hole Green function
\cite{Strinati} to connect the many-body wavefunctions and the
corresponding eigenenergies to the diagrammatic Green function
formulation of the last section. Thus, we may view the result of a
full configuration interaction calculation as the solution of a
generalized Bethe-Salpeter equation with the \emph{exact} two-particle
self-energy for the electrons of the localized
subsystem. Diagrammatically, this corresponds to

\begin{equation}
  \parbox{20mm}{\begin{fmfgraph}(20,15)
      \fmfleft{i1,i2} \fmfright{o1,o2}
      \fmf{plain,foreground=red}{i1,v2} \fmf{plain,foreground=red}{i2,v1}
      \fmfpoly{smooth,tension=0.5,filled=shaded}{v1,v2,v3,v4}
      \fmf{plain,foreground=red}{v3,o1} \fmf{plain,foreground=red}{v4,o2}
  \end{fmfgraph}}
  =
  \parbox{17mm}{\begin{fmfgraph}(17,15)
      \fmfleft{i1,i2} \fmfright{o1,o2}
      \fmf{plain,foreground=red}{i1,o1} \fmf{plain,foreground=red}{i2,o2}
  \end{fmfgraph}}
  +
  \parbox{17mm}{\begin{fmfgraph}(17,15)
      \fmfleft{i1,i2} \fmfright{o1,o2}
      \fmf{plain,foreground=red}{i1,o1} \fmf{plain,foreground=red}{i2,o2}
  \end{fmfgraph}}
  \left(
  \parbox{17mm}{\begin{fmfgraph}(17,15)
      \fmfleft{i1,i2} \fmfright{o1,o2}
      \fmf{plain,foreground=red}{i1,v1} \fmf{plain,foreground=red}{i2,v1}
      \fmf{plain,foreground=red}{v2,o1} \fmf{plain,foreground=red}{v2,o2}
      \fmf{dbl_dashes,foreground=blue}{v1,v2}
  \end{fmfgraph}}
  +
  \parbox{17mm}{\begin{fmfgraph}(17,15)
      \fmfleft{i1,i2} \fmfright{o1,o2}
      \fmf{plain,foreground=red}{i1,v1} \fmf{plain,foreground=red}{i2,v2}
      \fmf{plain,foreground=red}{v1,o1} \fmf{plain,foreground=red}{v2,o2}
      \fmf{dbl_dashes}{v1,v2}
  \end{fmfgraph}}
  +
  \parbox{20mm}{\begin{fmfgraph}(23,20)
      \fmfleft{i1,i2} \fmfright{o1,o2}
      \fmf{plain,foreground=red}{i1,v1} \fmf{plain,foreground=red}{i2,v2}
      \fmf{plain,foreground=red}{v1,v2}
      \fmf{plain,foreground=red}{v3,o1} \fmf{plain,foreground=red}{v4,o2}
      \fmf{plain,foreground=red}{v3,v4}
      \fmf{dbl_dashes,foreground=blue}{v1,v4}
      \fmf{dbl_dashes,foreground=blue}{v2,v3}
  \end{fmfgraph}}
  +...\right)
  \parbox{20mm}{\begin{fmfgraph}(20,15)
      \fmfleft{i1,i2} \fmfright{o1,o2}
      \fmf{plain,foreground=red}{i1,v2} \fmf{plain,foreground=red}{i2,v1}
      \fmfpoly{smooth,tension=0.5,filled=shaded}{v1,v2,v3,v4}
      \fmf{plain,foreground=red}{v3,o1} \fmf{plain,foreground=red}{v4,o2}
  \end{fmfgraph}},
\label{QCembed}
\end{equation}
where the term in brackets now contains more complicated irreducible
diagrams. Also, the fully screened interaction is not given by
Eq.~(\ref{RPAscreened}) any more, but contains more complicated
screening processes due to the exact treatment of the subsystem

\begin{equation}
\parbox{20mm}{\begin{fmfgraph}(20,20)
      \fmfleft{i1} \fmfright{o1}
      \fmf{dbl_dashes}{i1,o1}
  \end{fmfgraph}}
\;=\;
 \parbox{20mm}{\begin{fmfgraph}(20,20)
      \fmfleft{i1} \fmfright{o1}
      \fmf{dbl_dashes,foreground=blue}{i1,o1}
 \end{fmfgraph}}
\;+\;
\parbox{25mm}{\begin{fmfgraph}(25,20)
      \fmfleft{i1} \fmfright{o1}
      \fmf{dbl_dashes,foreground=blue}{i1,v1}
      \fmf{dbl_dashes,foreground=blue}{v2,o1}
      \fmf{plain,left,foreground=red,tension=0.7}{v1,v2,v1}
      \end{fmfgraph}}
\;+\;
\parbox{35mm}{\begin{fmfgraph}(35,20)
      \fmfleft{i1} \fmfright{o1}
      \fmf{dbl_dashes,foreground=blue}{i1,v1}
      \fmf{dbl_dashes,foreground=blue}{v2,v3}
      \fmf{dbl_dashes,foreground=blue}{v4,o1}
      \fmf{plain,left,foreground=red,tension=0.5}{v1,v2,v1}
      \fmf{plain,left,foreground=red,tension=0.5}{v3,v4,v3}
 \end{fmfgraph}}
\;+\;
\parbox{35mm}{\begin{fmfgraph}(35,20)
    \fmfleft{i1} \fmfright{o1}
    \fmftop{i3}  \fmfbottom{i4}
    \fmf{dbl_dashes,foreground=blue}{i1,v1}
    \fmf{dbl_dashes,foreground=blue}{v3,o1}
    \fmf{phantom,tension=1}{i3,v2}
    \fmf{phantom,tension=1}{i4,v4}
    \fmf{plain,left=0.4,foreground=red,tension=1}{v1,v2,v3,v4,v1}
    \fmf{dbl_dashes,foreground=blue,tension=0}{v2,v4}
\end{fmfgraph}}
\;+\; ...
\end{equation}

Again, the use of the environment-screened interaction in wavefunction
methods allows for the application of these methods to large systems
containing many electrons, which traditionally are out of reach for
these methods. In addition to the environment-screened interaction,
which only describes response properties of the environment, we note
that it is also necessary to include an additional external potential
into the quantum chemistry calculation, which is caused by the ground
state configuration of the environment.

\subsection{Self-Consistent Equations for Environment-Screened Potential}

Having demonstrated the usefulness of an environment-screened
interaction for excited state embedding approaches, we now turn to the
task of computing it. While the calculation of the fully screened
interaction is relatively straightforward in localized or periodic
systems, the calculation of the interaction with \emph{only}
environment screening proves more difficult if the localized system
constitutes an inhomogeneity in an otherwise homogeneous system, such
as a defect in a crystal. For this case, we demonstrate how matrix
elements of the environment-screened interaction can be calculated
from random phase approximation calculations of only the \emph{homogeneous}
system (crystal without defect) by subtracting out screening
contributions from the region where the inhomogeneity (defect)
resides. The desired matrix elements are computed from the
environment-screened \emph{potential} resulting from the ``external
charge density'' of a product of basis functions. This potential obeys
a self-consistent equation introduced in this section, which - in the
next section - is then transformed into a matrix form.

After choosing an explicit set of basis functions $g_i$ (assumed to be
real) for the quantum chemistry calculation of the embedded localized
subsystem, the interaction $v(\r,\rp)$ between electrons appears in
the resulting matrix form of Schroedinger's equation only in
interaction matrix elements of the form
\begin{equation}
\langle g_1, g_2 | v | g_3, g_4 \rangle =
\int d^3r \int d^3r' g_1(\r) g_2(\r) v(\r,\rp) g_3(\rp) g_4(\rp).
\label{MatrixElement}
\end{equation}

In our embedding approach, we replace the bare Coulomb potential
$\Phi(\r)=\int d^3r' g_3(\rp)g_4(\rp)/|\r-\rp|$ due to the ``charge
distribution'' $g_3 g_4$ in Eq.~(\ref{MatrixElement}) by the
environment-screened potential $\tilde{\Phi}$ according to
\begin{equation}
\langle g_1, g_2 | \Phi \rangle \rightarrow \langle g_1, g_2 |
\tilde{\Phi} \rangle.
\end{equation}

$\tilde{\Phi}$ takes into account the response of all environment
molecules to the ``external charge density'' $\rho_{ext}=g_3 g_4$ and
solves the self-consistent equation
\begin{equation}
\tilde{\Phi}= K (\rho_{ext} + \chi_{env}\tilde{\Phi}),
\label{SelfCons}
\end{equation}
where $K$ denotes the bare Coulomb operator $[K\rho](\r)=\int d^3r'
\rho(\rp)/|\r-\rp|$ and $\chi_{env}=\chi_{crys}-\chi_{sys}$ is the
environment polarizability given by the difference of the
polarizability of the full crystal (without the inhomogeneity) and the
polarizability of the local subsystem. We note that
$\chi_{env}\tilde{\Phi}$ is the self-consistent induced charge density
in the environment caused by the external charge $\rho_{ext}$.

Solution of Eq.~(\ref{SelfCons}) is complicated by the presence of
both quantities that describe extended systems and are best
represented in Fourier space, such $\chi_{crys}$, and quantities which
describe the explicit local system and are best represented by
localized functions, such as $\chi_{sys}$. We overcome this difficulty
by replacing solution of Eq.~(\ref{SelfCons}) by a two step process:
In the first step, we compute the effective field due to an external
charge $\rho_{tot}$ allowing the whole crystal (and not only the
environment region) to screen the external charge. Consequently,
$\tilde{\Phi}$ obeys
\begin{equation}
\tilde{\Phi}=K (\rho_{tot}+\chi_{crys}\tilde{\Phi}).
\label{Extended}
\end{equation}

Expressing all quantities in plane waves, this equation can be solved
straightforwardly yielding
\begin{equation}
\tilde{\Phi}=(K^{-1}-\chi_{crys})^{-1}\rho_{tot} \equiv K_{crys}\rho_{tot},
\label{SolveExtended}
\end{equation}
where $K_{crys}$ denotes the fully screened interaction of the
homogeneous system (e.g. perfect crystal without defect).

In the second step, we impose that the total external charge in
Eq.~(\ref{Extended}) must be given by
\begin{equation}
\rho_{tot}=\rho_{ext}+\Delta\rho = \rho_{ext} - \chi_{sys}\tilde{\Phi},
\label{Explicit}
\end{equation}
which involves only the localized system response and can be solved in
a localized representation.  Inserting Eq.~(\ref{Explicit}) into
Eq.~(\ref{Extended}) clearly gives back
Eq.~(\ref{SelfCons}). Alternatively, we can substitute
Eq.~(\ref{SolveExtended}) into Eq.~(\ref{Explicit}) yielding
\begin{equation}
\Delta\rho = -\chi_{sys}K_{crys}\rho_{tot}=-\chi_{sys}K_{crys}(\rho_{ext}+\Delta\rho),
\label{MainEquation}
\end{equation}
which now constitutes a self-consistent equation for $\Delta \rho$. In
the following, we will make use of this latter equation.

\subsection{Transformation to matrix equation}

We now show how to transform Eq.~(\ref{MainEquation}) for the charge
density $\Delta \rho$ induced in the environment into a matrix
equation.

We employ the random-phase approximation for the polarizability of the
explicit system
\begin{equation}
\chi_{sys}(\r,\rp|\omega) = 2 \sum_{jk} (f_k-f_j)
\frac{ \phi^*_k(\r) \phi_j(\r) [\phi^*_k(\rp) \phi_j(\rp)]^*}
{\omega - (\epsilon_j-\epsilon_k) + i\eta},
\label{PolCrys}
\end{equation}
where $f_i=0$ or $1$ are occupation factors, $\phi_i(\r)$ and
$\epsilon_i$ denote single-particle wavefunctions and orbital energies
of the local system, respectively.  From now on, we will work in the
static approximation setting $\omega=0$, which is well-justified for
many applications \cite{DelSole}.

Introducing the standard transition-space notation, where the
transition of an electron from orbital $k$ to orbital $j$ is labeled
by $\mu=(k,j)$, we express $\chi_{sys}$ as
\begin{equation}
\chi_{sys}(\r,\rp)=\sum_{\mu=(v,c)}\rho_\mu(\r)\chi_\mu\rho^*_\mu(\rp),
\end{equation}
with $\chi_\mu=4/(\epsilon_j-\epsilon_k)$,
$\rho_{\mu}(\r)=\phi^*_k(\r)\phi_j(\r)$ and the indices $v$ and $c$
run over occupied and empty orbitals, respectively. Inserting this
expression for $\chi_{sys}$ into Eq.~(\ref{MainEquation}) yields
\begin{equation}
\Delta\rho(\r) = -\sum_{\mu} \rho_\mu(\r) \chi_\mu \int d^3r' \int d^3r''
\rho^*_\mu (\rp) K_{crys}(\rp,\rpp) [ \rho_{ext}(\rpp) + \Delta\rho(\rpp) ].
\end{equation}

The ansatz $\Delta\rho(\r) = \sum_\mu A_\mu \rho_\mu(\r)$ then leads
to the following matrix equation for the coefficients $A_\mu$
\begin{equation}
A_\mu= - \chi_\mu \left( a_\mu + \sum_\nu M_{\mu\nu}A_\nu \right),
\label{A}
\end{equation}
where we defined
\begin{eqnarray}
a_\mu &=& \int d^3r \int d^3r' \rho^*_\mu(\r)K_{crys}(\r,\rp)\rho_{ext}(\rp), \\
M_{\mu\nu} &=& \int d^3r \int d^3r' \rho^*_\mu(\r)K_{crys}(\r,\rp)\rho_\nu(\rp).
\end{eqnarray}

Solving Eq.~(\ref{A}) we arrive at
\begin{equation}
A = -[\chi^{-1}+M]^{-1}a,
\label{SolveA}
\end{equation}
with $\chi^{-1}_{\mu\nu}=\chi^{-1}_{\mu}\delta_{\mu\nu}$. In sum, we
solved Eq.~(\ref{MainEquation}) for $\Delta \rho$ in terms of the
transition-space matrix $M$ and the vector $a$, which involve the
fully screened interaction $K_{crys}$ of the corresponding homogeneous
system. In the next section, we describe how these quantities are
computed within the random-phase approximation.

\subsection{Calculation of $M$ and $a$}
 
Exploiting the translational symmetry of the crystal without the
inhomogeneity (i.e. the localized subsystem), we express $K_{crys}$,
$\rho_{ext}$ and $\rho_\mu$ as integrals over the Brillouin zone. For
example, $\rho_\mu$ is given by
\begin{equation}
\rho_\mu(\r) = \int_{BZ} \frac{d^3q}{V_{BZ}} \rho_{\mu,\q}(\r) e^{i\q\cdot\r},
\end{equation}
where $V_{BZ}=(2\pi)^3/V_{cell}$ denotes the volume of the Brillouin
zone with $V_{cell}$ being the unit cell volume. We note that
$\rho_{\mu,\q}(\r)=\sum_\G \rho_{\mu,\q}(\G) \exp(i\G\cdot\r)$, where
$\G$ denotes a reciprocal lattice vector, is a lattice-periodic
function.

To compute $K_{crys}$ [Eq.~(\ref{SolveExtended})], the polarizability
of the crystal is taken in the static random-phase approximation
\cite{Adler,Wiser} and explicitly given by
\begin{equation}
\chi_{crys,\q}(\G,\Gp)=\frac{4}{N_qV_{cell}} \sum_{cv\k}
\frac{\langle v,\k|e^{-i(\q+\G)\cdot\r}|c,\k+\q\rangle
\langle c,\k+\q|e^{i(\q+\Gp)\cdot\rp}|v,\k\rangle}
{\epsilon_{v,\k}-\epsilon_{c,\k+\q}},
\label{chiFull}
\end{equation}
where $N_q$ is the number of sample points in the Brillouin zone and
the indices $c$ and $v$ run over conduction and valence states,
respectively. In the molecular crystals we are interested in orbitals
on neighboring molecules overlap very little, allowing for additional
simplifications: (i) we assume flat bands,
i.e. $\epsilon_{n,\k}=\epsilon_{n}$, and (ii) we assume that the
crystalline Bloch states can be constructed from the wavefunction at
the $\Gamma$-point through a tight-binding procedure via
\begin{equation}
\phi_{n\k}(\r)=\frac{1}{\sqrt{N_q}}\sum_\R e^{i\k\cdot\R} w_n(\r-\R),
\label{WannierTrans}
\end{equation}
where $\R$ runs over all $N_q$ units cells comprising the crystal and
$w_n(\r)=\psi_{n\k=0}(\r)$ in the unit cell surrounding the origin and
$w_n(\r)=0$ in all other cells. Formally, Eq.~(\ref{WannierTrans})
describes the Wannier transformation from localized orbitals or
Wannier functions to Bloch states using an approximate Wannier
function $w_n$ [note that the exact Wannier function is given by
$w_{n\R}=1/\sqrt{N_q} \times \sum_\k \exp(-i\k\cdot\R)\psi_{n\k}$].

The tight-binding ansatz eliminates the summation over the Brillouin
zone in Eq.~(\ref{chiFull}) and yields
\begin{equation}
\chi_{crys,\q}(\G,\Gp)=\frac{4}{V_{cell}}\sum_{vc}\langle v|e^{-i(\q+\G)\cdot\r}|c\rangle
\frac{1}{\epsilon_v-\epsilon_c} \langle c|e^{i(\q+\Gp)\cdot\rp}|v\rangle,
\end{equation}
where $ \langle v|e^{-i(\q+\G)\cdot\r}|c\rangle = \int_{V_{cell}}d^3r
\phi^*_v(\r) e^{-i(\q+\G)\cdot\r} \phi_c(\r)$. We point out that the
tight-binding approximation for the crystal polarizability is invoked
only for numerical convenience and that all following expressions can
also be evaluated using the full RPA polarizability, given in
Eq.~(\ref{chiFull}), instead.

Finally, we express $K_{crys}$ in terms of the symmetrized dielectric
matrix
$\tilde{\epsilon}_\q(\G,\Gp)=\delta_{\G\Gp}-K^{1/2}_\q(\G)\chi_{crys,\q}(\G,\Gp)K^{1/2}_\q(\Gp)$
via
\begin{equation}
K_{crys,\q}(\G,\Gp)=K^{1/2}_\q(\G)\tilde{\epsilon}^{-1}_\q(\G,\Gp)K^{1/2}_\q(\Gp),
\end{equation}
with $K_\q(\G)=4\pi/|\q+\G|^2$. The components of $M$ and $a$ can now be
expressed completely in Fourier space and are given by
\begin{eqnarray}
a_{\mu}=V_{cell}\int \frac{d^3q}{V_{BZ}} \sum_{\G\Gp}
\rho^*_{\mu,\q}(\G) K_{crys,\q}(\G,\Gp) \rho_{ext,\q}(\Gp), \label{a}\\
M_{\mu\nu}=V_{cell} \int \frac{d^3q}{V_{BZ}} \sum_{\G\Gp}
\rho^*_{\mu,\q}(\G) K_{crys,\q}(\G,\Gp) \rho_{\nu,\q}(\Gp). \label{M}
\end{eqnarray}

\subsection{Environment-screened interaction matrix element}

The final expression for the environment-screened matrix element is
\begin{eqnarray}
\langle g_1, g_2|\tilde{\Phi} \rangle &=&
\langle g_1, g_2| K_{crys}|\rho_{ext}(3,4) + \Delta \rho(3,4)\rangle \nonumber \\
&=& \langle g_1, g_2| K_{crys}|g_3, g_4\rangle +
\langle g_1, g_2|K_{crys}|\sum_\mu A_{\mu}(3,4) \rho_\mu\rangle \nonumber \\
&=& \langle g_1, g_2| K_{crys}|g_3, g_4\rangle - a^\dagger(1,2) X a(3,4),
\label{MatrixElementFinal}
\end{eqnarray}
where we defined $X=[\chi^{-1}+M]^{-1}$ and we used
Eqs.~(\ref{SolveExtended}), (\ref{Explicit}) and (\ref{SolveA}). We
also now explicitly write $a(3,4)$ to denote explicitly the dependence
on the basis functions $g_i$ (similarly for $\Delta\rho$ and $A$).

\subsection{Numerical challenges}

In this section, we discuss the two major numerical difficulties of
the described embedding approach: (i) the integrations over the
Brillouin zone in Eqs.~({\ref{a}}) and (\ref{M}) and (ii) the handling
of extremely narrow basis functions.

Regarding the first issue, we note that the Brillouin zone integrals
in Eqs.~({\ref{a}}) and (\ref{M}) must be handled with great care,
because certain elements of $K_{crys,\q}$ diverge as $|\q|$ approaches
zero: the head $K_{crys,\q}(0,0)$ diverges as $1/\q^2$ and the wings,
$K_{crys,\q}(0,\Gp)$ and $K_{crys,\q}(\G,0)$, diverge as $1/|\q|$.  Although
these singularities are integrable, they can cause extremely slow
convergence when the integral is approximated by a discrete sum over
sample points in the Brillouin zone.
 
Most approaches \cite{Ambrosch,ShamSchluterGodby} use a regular mesh
including the origin to carry out the Brillouin zone integration. The
contribution from the volume element at the origin is obtained by
approximating the smooth part of the integrand by its value at the
origin, replacing the volume element by a sphere of equal volume and
finally analytically integrating the divergent part of the integrand
in the spherical volume. While numerically efficient, we expect the
error associated with this scheme to converge quite slowly. Also, in
anisotropic systems the smooth part of the integrand might not even be
well-defined at the origin and instead vary strongly depending on the
direction of approach towards the origin.
 
In contrast, in our approach we first isolate the singular parts of
the integrand by splitting the integral over the Brillouin zone into
two parts. The part containing the singularity is transformed to
spherical coordinates thus removing the singularity, while the other
part is evaluated using a regular grid over the Brillouin zone. For a
generic function $f(\q)$, we write
\begin{equation}
\int d^3q f(\q) = \int d^3q (1-e^{-\sigma^4 \q^4}) f(\q) + \int d^3q e^{-\sigma^4 \q^4} f(\q),
\end{equation}
where $\sigma$ is a parameter which can be adjusted to optimize
convergence. Note that the integrand in the first term on the right
hand side is well-behaved at the origin, even if $f(\q)$ diverges as
$1/\q^2$. We use a simple regular grid over the whole Brillouin zone to
evaluate this term. The integrand of the second term is divergent at
the origin if $f(\q)$ diverges, but vanishes rapidly as $|\q|$ increases
because of the fast decay of $\exp(-\sigma^4\q^4)$. The singularity can
be removed by transforming the integral to spherical coordinates
yielding
\begin{equation}
\int d^3q e^{-\sigma^4 \q^4} f(\q) = \int dq \left( q^2 e^{-\sigma^4 q^4} \int
d\Omega f(q,\Omega) \right),
\label{EwaldTrick}
\end{equation}
where the integrand in parentheses is now well-behaved at the origin
and relatively smooth, such that we can use a Gaussian quadrature
scheme for accurate evaluation. We also use Gaussian quadrature to
carry out the integral over spherical angles.

Our scheme thus avoids the approximations described above. In
particular, the contribution from the singular region surrounding the
origin is computed with high accuracy. Also, the smooth part of the
integrand is never evaluated at the origin, where it is not
necessarily uniquely defined, making our scheme ideal for the study of
anisotropic systems.

Regarding the second issue raised above, the difficulty of handling
extremely narrow basis functions, we point out that, in actual
density-functional calculations, the Kohn-Sham orbitals are
represented on discrete real-space or Fourier space grids, which are
typically too coarse to faithfully represent the narrow basis
functions used in quantum chemistry calculations.  However, because
the crystal polarizability [Eq.~(\ref{chiFull})] entering the fully
screened interaction is computed from Kohn-Sham orbitals only Fourier
components of the ``external charge density'' (caused by a pair of
basis functions) which belong to the grid of the density-functional
calculation are screened, suggesting that the crystal-screened
interaction can be decomposed as
\begin{equation}
K_{crys}=K+\Delta K_{crys},
\end{equation}
where $\Delta K_{crys}$ now only acts on the space of functions
representable on the grid of the density-functional
calculation. Instead of the full environment-screened interaction
matrix element [Eq.~(\ref{MatrixElementFinal})], the calculation of
which would require grids fine enough to represent the narrow basis
functions, we compute the difference $\langle g_1 g_2|\Delta
K_{crys}|g_3 g_4 \rangle$, which allows us to work with the grids of
the density-functional calculation only. This difference has to be
added to the bare Coulomb matrix element, which is computed with high
accuracy in existing quantum chemistry codes.

\subsection{External potential due to environment}

In addition to the environment-screened interaction, we also include a
static external potential caused by the ground state configuration of
the environment into the quantum chemistry calculation. The inclusion
of this additional external potential is necessary, because the
environment-screened interaction only accounts for \emph{response
  effects} in the environment, but not for the \emph{static effect of
  the environment ground state configuration} on the electrons of the
localized subsystem. The proper definition of the external potential
requires great care in order to avoid ``double counting'' of
interaction processes between the explicit system and the environment.

To reproduce the correct change in electron density $\delta \rho$ in
the localized system system upon embedding, the external potential
$\phi_{ext}$ must fulfill
\begin{equation}
\delta \rho = \chi_{scr} \, \phi_{ext},
\label{PhiExt}
\end{equation}
where $\chi_{scr}$ denotes the response function of the localized
subsystem with an environment-screened interaction between the
electrons and we assumed that $\phi_{ext}$ is relatively weak.  If the
wavefunction overlap between subsystem and environment is small, we
can relate $\phi_{ext}$ to the total electrostatic potential
$\phi_{env}$ due to the environment ground state
configuration. $\phi_{env}$ is given by
\begin{equation}
\phi_{env} = K \rho_{env},
\label{staticPot}
\end{equation}
where the environment charge density $\rho_{env}$ is the sum of
electronic and nuclear contributions. From its definition,
$\phi_{env}$ produces the correct change in density $\delta \rho$,
when applied to the subsystem with \emph{unscreened interactions}
between the electrons. If the potential is weak, this implies
\begin{equation}
\delta \rho = \chi_0 \, \phi_{env},
\label{PhiEs}
\end{equation}
where $\chi_0$ denotes the response function of the subsystem with
unscreened interactions. Combining Eq.~(\ref{PhiExt}) and
Eq.~(\ref{PhiEs}), we find
\begin{equation}
\phi_{ext}=\chi^{-1}_{scr} \, \chi_0 \, \phi_{env},
\end{equation}
which allows for the calculation of the correct external potential
from the electrostatic potential caused by the environment ground
state configuration.

\section{Application to organic crystals}

In this section, we apply the described embedding approach to study
electronic excitations in crystals consisting of short organic
molecules. In these crystals, we take the molecules of a single unit
cell as the localized subsystem and all other molecules as the
environment.

As a test case, we investigate crystalline ethylene.  Crystalline
ethylene being one of the simplest organic molecular crystals has been
studied for a long time\cite{bunn,dows,taddei}.

The excitations of isolated short organic molecules have attracted
considerable attention because of the non-trivial character of their
low-lying excited states \cite{HeadGordon,Kohler,McDiarmid}. For
example, contrary to naive expectations the lowest excited state of
the octatetraene molecule is not a one-electron (HOMO-LUMO)
excitation, but has a HOMO$^2$-LUMO$^2$ double-excitation character
\cite{Kohler, McDiarmid}.

Upon embedding the molecule in a crystalline environment, we expect
the molecular excitations to be altered because of (i) the screening
by the crystalline environment and (ii) the delocalization of the
excited state over several neighboring molecules. For crystals of
short acenes, Ambrosch-Draxl and coworkers \cite{Ambrosch2} found that
there is very little delocalization of the excitation. We therefore
only include the two inequivalent molecules of a single unit cell into
our embedded quantum-chemistry calculation.

\subsection{Computational details}

We carry out density-functional calculations of ethylene crystals
using the generalized gradient approximation \cite{pbe}.  We employ
Kleinman-Bylander pseudopotentials \cite{Kleinman}, a plane wave
cutoff of $40$ Hartree and a $2\times 2\times 2$ Monckhorst-Pack
\cite{Monckhorst} kpoint grid. For ethylene, the crystal structure is
well-known from x-ray and neutron diffraction and we employ the
lattice parameters and atomic positions provided by
Ref.~\cite{ElliotLeroi} in our calculation.  We then relax the ionic
positions and compute a large number of empty bands, which are needed
for the calculation of the polarizability [Eq.~(\ref{PolCrys})].

Next, we evaluate the environment-screened interaction matrix
elements, Eq.~(\ref{M}). In this calculation, we only employ a subset
of the reciprocal lattice used in the density-functional calculations
consisting of $\sim 1000$ reciprocal lattice vectors. We find that the
matrix elements are sufficiently converged if $200$ empty bands are
used. For the Brillouin zone summation, we employ a regular grid
containing $512$ kpoints for the first integral in
Eq.~(\ref{EwaldTrick}) and $270$ kpoints for the second integral,
which contains the singularity.  Finally, we approximate the external
potential by the electrostatic potential of the environment,
$\phi_{ext} \approx \phi_{env}$.

To compute excited states of the localized subsystem, we carry out
configuration interaction calculations using the 6-31G basis set.

\subsection{Results}

\begin{table}
  \setlength{\doublerulesep}{0\doublerulesep}
  \setlength{\tabcolsep}{5\tabcolsep}
  \begin{tabular}{c | c c}
    \hline\hline\\
            & bare         & screened\\
            & [eV]         & [eV]        \\
    \hline
    state 1  & 17.432  &  17.384  \\
    state 2  & 17.436  &  17.554 \\
    state 3  & 19.004  &  19.004 \\
    \hline\hline
  \end{tabular}
  \caption{Lowest excitations energies (measured from the ground state
    energy) in crystalline ethylene. The first column contains
    excitation energies obtained from configuration interaction
    calculations of the two molecules in a unit cell without any
    environment effects. The second column contains excitation
    energies for the unit cell embedded in the crystalline
    environment.}
  \label{ethylene}
\end{table}

Table~\ref{ethylene} shows our results for the energies of lowest
lying excitations in crystalline ethylene. We observe that the changes
due to the environment screening are smaller than one
percent. Interestingly, we find that the inclusion of environment
screening lowers the energy of the lowest excited state, but increases
the energy of the second lowest excited state. The energy of the third
lowest excited state remains constant.

\subsection{Discussion and Outlook}
In this work, we have described a general framework for embedding
theories of excited states in physical systems where the excitation
is localized on a small subsystem. Contrary to previous approaches, we
fully include the self-consistent screening response of the
environment by using an effective interaction between electrons in the
explicit subsystem. Once obtained, the environment-screened
interaction may be employed in various highly accurate electronic
structure description of the localized subsystem: we demonstrate how
environment effects can be incorporated into Green function methods or
wavefunction methods, which are typically employed in quantum
chemistry.

We apply our embedding theory to the calculation of excited state
energies of crystalline ethylene and find encouraging results.

\end{fmffile}
\bibliography{Sum}
\end{document}